\documentclass[11pt]{article}
\begin{document}
\title{High-Fidelity Teleportation of Independent Qubits}
\author{Dik Bouwmeester$^1$, Jian-Wei Pan$^2$, Harald
Weinfurter$^3$,\\ and Anton Zeilinger$^2$}
\maketitle
\begin{center}
 $^1$
\em{Clarendon Laboratory, Oxford University, Parks Road, Oxford
OX1 3UP, United~Kingdom} \\ $^2$ \em{{\it Institut f\"{u}r
Experimentalphysik, University of Vienna, Boltzmanngasse 5,
1090~Wien, Austria}} \\ $^3$ \em{Sektion
Physik,Ludwig-Maximilians-Universit\"{a}t M\"{u}nchen,
Schellingstr. 4/III, D-80799~M\"{u}nchen, Germany}
\end{center}

\begin{abstract}
Quantum teleportation is one of the essential primitives of
quantum communication. We suggest that any quantum teleportation
scheme can be characterized by its efficiency, i.e. how often it
succeeds to teleport, its fidelity, i.e. how well the input state
is reproduced at the output, and by its insensitivity to cross
talk, i.e. how well it rejects an input state that is not intended
to teleport. We discuss these criteria for the two teleportation
experiments of independent qubits which have been performed thus
far. In the first experiment (Nature {\bf 390},575 (1997)) where
the qubit states were various different polarization states of
photons, the fidelity of teleportation was as high as 0.80 $\pm$
0.05 thus clearly surpassing the limit of 2/3 which can, in
principle, be obtained by a direct measurement on the qubit and
classical communication. This high fidelity is confirmed in our
second experiment (Phys. Rev. Lett. {\bf 80}, 3891 (1998)),
demonstrating entanglement swapping, that is, realizing the
teleportation of a qubit which itself is still entangled to
another one. This experiment is the only one up to date that
demonstrates the teleportation of a genuine unknown quantum state.
\end{abstract}

\section{Introduction}
Two of the most fundamental protocols of quantum communication are
quantum teleportation and entanglement swapping
\cite{BEN93,ZUK93}, the teleportation of an entangled state. With
the qubit being the elementary representative of information in
the quantum domain, teleportation and entanglement swapping of
qubits are essential contributions to any quantum communication
toolbox. Thus far there have been two experiments performed
\cite{BOU97,PAN98} on the teleportation of independent qubits.
Another experiment \cite{ROM99} demonstrated the quantum
teleportation protocol not for an independent qubit but for a
qubit that has to be prepared on a specific particle (entangled
with another particle). And finally, a fourth experiment
\cite{FUR98} demonstrated the quantum teleportation for continuous
variables. In the present paper we suggest ways how to
characterize the quality of a given teleportation scheme and we
discuss specifically the experiments on teleportation of
independent qubits \cite{BOU97,PAN98} from that perspective. We
show explicitly that it is important to distinguish teleportation
fidelity from teleportation efficiency. That way some criticism
which has been raised in the literature \cite{COM1} turns out to
be unjustified \cite{COM2}. In section \ref{setting} we will first
briefly review the two experiments concerning teleportation of
independent qubits before giving our criteria for experimental
quantum teleportation in section~\ref{criteria}. In sections
\ref{single} and \ref{entangle} the two experiments will be
analyzed in view of the given criteria. Conclusions are drawn in
section \ref{conc}.

\section{Experimental Quantum Teleportation of Independent Qubits}
\label{setting} In the quantum teleportation experiment presented
in Ref.~\cite{BOU97} an incoming UV pump-pulse has two
opportunities to create pairs of photons (Fig.1). The idea is that
on the path from left to right the pulse creates an entangled
pair. This is the ancillary entangled pair of the original
proposal \cite{BEN93}. One of the ancillaries is passed on to
Alice and the other one to Bob. The latter one will obtain the
teleported qubit encoded in its polarization. On the return path
the pulse again creates a pair of photons where in the original
experimental teleportation scheme the fact that the two are
entangled was not utilized. In fact, one of these two photons was
passed through an adjustable polarizer such defining the state
(the initial qubit) to be teleported. This procedure breaks the
entanglement for that pair. The second photon of that pair is sent
to a trigger detector whose purpose it was to reject all detector
events where this second pair was not created. In the experiment
the entangled photons, photons 2 and 3 in Fig.1, were produced in
the anti-symmetric state
\begin{equation}
|\Psi^{-}\rangle_{23}= \frac{1}{\sqrt{2}} \left( | \mbox{H}
\rangle_2 | \mbox{V} \rangle_3 - | \mbox{V} \rangle_2 | \mbox{H}
\rangle_3 \right) \,,
\end{equation}
where $| \mbox{H} \rangle$ and  $| \mbox{V} \rangle$ represent the
horizontally- and vertically-polarized photon state.

The idea of the experiment then is that Alice subjects the photon
to be teleported and her ancillary photon to a (partial)
Bell-state measurement using a beam-splitter. Observation of a
coincidence at the Bell-state analyzer detectors f1 and f2 then
informs Alice that her two photons were projected into the
anti-symmetric state $|\Psi^{-}\rangle_{12}$.

This then implies that Bob's photon is projected by Alice's
Bell-state measurement onto the original state. This can be seen
by assuming that it is the intention of the experiment to be able
to teleport the general qubit,
\begin{equation}
\label{gen}
|\Psi\rangle_{1}= \alpha | \mbox{H} \rangle_1 + \beta
| \mbox{V} \rangle_1  \,,
\end{equation}
with $\alpha$ and $\beta$ complex amplitudes satisfying
$|\alpha|^2+|\beta|^2=1$. Then the initial state of qubit plus
ancillaries is given by the product state
\begin{equation}
|\Psi\rangle_{123}= |\Psi\rangle_{1}|\Psi^{-}\rangle_{23}\,.
\end{equation}
Projection of photon 1 and 2 onto the anti-symmetric state yields
\begin{equation}
\langle \Psi^{-}|_{12}| \Psi_{123} \rangle= -\frac{1}{2} \left(
\alpha | \mbox{H} \rangle_3 + \beta | \mbox{V} \rangle_3 \right)
\,.
\end{equation}
This indicates that the polarization state determined by the
complex amplitude $\alpha$ and $\beta$ has been transferred from
photon 1 to photon 3.  The amplitude factor of 1/2 indicates that
only in one out of four cases the result of the Bell-state
measurement is the anti-symmetric one \cite{BEN93,BOU97}.

The experimental scheme then simply proceeded in defining various
different polarization states using polarizers and wave plates and
verifying by polarization measurement that Bob's photon actually
had the state adjusted by the polarizers and wave-plates it never
saw given that the coincidence between the detectors f1-f2 did
indicate a ($|\Psi^{-}\rangle_{12}$) Bell-state measurement. In
order to demonstrate the generality of the scheme it is not enough
to just demonstrate the teleportation of the base states $|H
\rangle$ and $|V \rangle$, which readily succeeded in experiment,
but also to demonstrate superpositions of these states. In the
experiment it was decided to demonstrate teleportation both for
two real-coefficients superpositions (linear polarization),  and
for one superposition with imaginary-coefficients representing
circular polarization.

The second experiment \cite{PAN98} demonstrated the teleportation
of an entangled state \cite{BEN93} by verifying the protocol of
entanglement swapping \cite{ZUK93}. Experimentally, the essential
difference was that in that experiment (Fig.2) the entanglement of
the pair created by the pulse upon its return passage was also
fully utilized. Therefore, in that experiment there was no
polarizer in the path of that photon of the second pair which was
sent to Alice's Bell-state analyzer thus not breaking the initial
entanglement. This means that the state when two separate pairs
were created in the way described reads
\begin{equation}
|\Psi\rangle_{1234}=| \Psi^{-}\rangle_{14}|\Psi^{-}\rangle_{23}\,,
\end{equation}
which is a product state of two entangled pairs. Observation of a
coincidence at the detectors f1-f2 again indicates that photon 1
and 2 have been projected into the anti-symmetric Bell-state which
now indicates that the final state is $|\Psi^{-}\rangle_{34}$.
This shows that now the outer two photons 3 and 4 have become
entangled. This can be seen as teleportation either of the state
of photon 2 over to photon 4 or the state of photon 1 over to
photon 3. Those viewpoints are completely equivalent. The
remarkable feature of that experiment is that the actually
teleported state is a photon state which is not well defined.
Since, as is well known, the state of a particle which is
maximally entangled to another one has to be described by a
maximally mixed density matrix. Indeed, in that experiment neither
of the two photons subject to the Bell-state measurement enjoyed a
quantum state on its own. They were both maximally mixed.
Therefore, what is teleported in such a situation is not the
quantum state of the photon but just the way how it relates to the
other photon it has been entangled to initially.

In order to demonstrate that teleportation succeeds in that case,
it is necessary to show that photons 3 and 4 are now entangled
with each other. This can been done by showing that the
polarizations of the two photons are always orthogonal
irrespective of the detection basis chosen \cite{PAN98}.

\section{Criteria for Experimental Quantum Teleportation}
\label{criteria} We will now identify criteria and notions by
which the quality of a certain teleportation procedure can be
evaluated. This may also serve to a certain extent as a means for
comparison of different teleportation procedures. However, it will
turn out that it seems impossible to define just a single
parameter which would serve to characterize all procedures.

Any quantum teleportation procedure can be characterized by how well it
can answer the following questions:
\begin{itemize}
\item[1]{How well can it teleport any arbitrary quantum state it is intended
to teleport? This is the fidelity of teleportation.}
\item[2]{How often does it succeed to teleport, when it is given an input
state within the set of states it is designed to teleport? This is the
efficiency of teleportation.}
\item[3]{If given a state the scheme is not intended to teleport, how well
does it reject such a state? This is the cross-talk rejection
efficiency.}
\end{itemize}

But foremost, one has to define the set of states the
teleportation procedure should be able to handle. It is of little
use to talk about a specific procedure but use the wrong states to
characterize its performance. The aim of the experiments presented
in Refs. \cite{BOU97,PAN98} (Innsbruck experiments) has been to
teleport with high fidelity a qubit, i.e. a two-dimensional
quantum state, given by the polarization state of a single photon.
Experiments performed at Caltech addressed the transfer of an
infinite dimensional quantum state represented by the continuous
quadrature amplitude components of an electro-magnetic field
\cite{FUR98}. We want to emphasize, that if one talks about one or
the other type of experiments one should use the appropriate
states for describing it.

In the following two sections we evaluate the above criteria in
detail for the two teleportation schemes realized in Innsbruck,
particularly in view of the criticism initially voiced by
Braunstein and Kimble \cite{COM1,KOK,RAL}. It is explicitly not
our intention to criticize the Caltech experiment though it will
be obvious from our analysis that the claim voiced by Kimble a
number of times that the Caltech experiment is the first bona fide
verification of quantum teleportation is unjustified.

\section{Teleportation of Single Qubits}
\label{single} Let us now analyze the first Innsbruck
teleportation experiment of independent qubits \cite{BOU97}. Since
it is the intention of the experiment to be able to teleport the
general qubit (Eq.~\ref{gen}) encoded in the polarization state of
a single photon, it is require (a) that the scheme is able to
teleport any superposition of this form with high fidelity and (b)
that the scheme does not teleport anything which is not of this
form.

What happens, if the system does not output a single photon
carrying the desired qubit? This situation can be treated on the
same footing as some absorption process along a communication
channel. As it is well known from other applications of
single-photon quantum communication, like quantum cryptography or
quantum dense coding, this will influence the efficiency of the
communication system but does not influence the coherence
properties of the remaining photons. This comes from the fact that
the possibility of absorption of the single photon, the "carrier"
of the qubit, does not alter the qubit itself. After
renormalisation of a two-dimensional state, the original state,
the qubit, is obtained again without any influence on the
teleportation fidelity. The situation changes drastically if one
considers the Caltech teleportation experiment of the quadrature
amplitude components of an electro-magnetic field. In that case,
an absorption of light-quanta changes the amplitudes of the
various Fock-states and therefore unavoidably changes the quantum
state that is transmitted. Consequently, absorption necessarily
decreases the fidelity of the teleportation procedure for
continuous variables but not for single qubits.

As explained in section 2 an incoming UV pump pulse has two
opportunities to create pairs of photons (Fig. 1). This can happen
either on the path from left to right or on the return path. The
cases where only one pair is produced can be rejected since only
the situations are accepted in which  the trigger detector p fires
together with both Bell-state analyzer detectors f1 and f2. Also,
any cases where more than two pairs are created can safely be
ignored because in the experiment the total probability of
creating one pair per pulse in the modes actually detected is of
the order 10$^{-4}$, which gives a detection rate of three pairs
from a single pump pulse of much less than one per day at the
experimental parameters.

What then does a three-fold coincidence p-f1-f2 tell us? There are
two possibilities. One is that we actually had a case of
teleportation of the initial qubit encoded in photon 1. In the
experiment this was demonstrated for the 5 polarizer settings H,
V, +45$^{\circ}$, -45$^{\circ}$ and R (circular). These settings
represent non-orthogonal qubits and altogether cover very
different directions on the Poincare sphere what provides a proof
that the scheme works for an arbitrary superposition. H and V
proved the working of the scheme for the natural basis states
defined by the properties of the experimental setup. The
+45$^{\circ}$ and - 45$^{\circ}$ linear polarization states proved
the proper operation for coherent superpositions with real
probability amplitudes and the R state for imaginary amplitudes.
This is sufficient to demonstrate that the scheme will work for
any superposition, in contrast to the suggestion by  Vaidman
\cite{VAID} that more settings are required for a full proof. The
fact that the transfer of a quantum state worked for
non-orthogonal states is a direct indication that entanglement is
at the heart of the experiments.

The second case when a p-f1-f2 coincidence can occur is when both
photon pairs are created by the pulse on its return trip. Thus, in
that case no teleported photon arrives at Bob's station  and
teleportation did not happen. Yet Alice recorded a coincidence
count at her Bell state detector. It has been argued by Braunstein
and Kimble that this possibility reduces the fidelity of our
teleportation scheme. Yet, as we will show now, it actually is an
advantage of our scheme that teleportation did not occur in that
case. Indeed, the state behind the polarizer in that case contains
two identically prepared photons. Therefore, since according to
our protocol, we only wish to teleport qubits encoded in
single-photon states, it is an advantage of our scheme that
teleportation does not occur. Thus our scheme has a high intrinsic
cross-talk rejection efficiency for these cases.

It might be argued that a spurious coincidence trigger at Alice's
Bell state analyzer reduces the usefulness of such a teleportation
scheme. Yet, all that happens is that Bob in such a case does not
receive a teleported photon even as the message he receives from
Alice might indicate that. There is no problem with that since he
was not supposed to have obtained a teleported photon in that case
anyway, as the state given to Alice does not fall within the class
of states, namely single-photon qubits,  the scheme is intended to
work for. That Alice falsely thinks that teleportation worked in
that case does not do any harm.

Another problem already discussed in the original publication
\cite{BOU97} is the fact that only one of the four Bell states was
identified. This simply means that the procedure works in 25\% of
the situations. Only whenever the state the two photons at the
Bell state analyzer were projected into happened to be the anti
symmetric one it was identified by a coincidence behind the beam
splitter. In the other 75\% of the cases teleportation was not
performed. Which of the Bell-states is actually observed is
independent of the qubit given to Alice! All this means is simply
that the efficiency of the scheme was significantly reduced
without any influence on the fidelity of the qubit quantum
teleportation.

Losses occur anyway in any realistic scheme and it is always
necessary in any protocol to provide for these cases by means of
some communication between the various participants. In that
spirit we emphatically stress that a reduction of the efficiency
of the procedure, of the fraction of cases where it actually
finished, does not reduce at all the fidelity which describes how
well the actually teleported qubit agrees with the original one.

Clearly, even if a teleportation procedure is inefficient in the
sense of rarely teleporting the given qubit, the fidelity of those
qubits which are teleported can be very high. This is very
different from a scheme which finishes the teleportation procedure
very frequently but with low fidelity. We will see below that the
experiment on entanglement swapping provides a clear case where
the distinction between efficiency and fidelity is obvious.

As evidenced by the final verification of the teleported qubits,
that is by a polarization measurement, the measured qubit
teleportation fidelity was rather high in the experiment
\cite{BOU97}. As can be seen from Fig. 3, it typically was of the
order of 0.80. This very clearly surpasses the limit of 2/3
indicated by the dotted line which at best could have been
obtained by Alice performing a polarization measurement on the
given photon, informing Bob about the measurement result via
classical communication, and by Bob accordingly preparing a photon
at his output.

In conclusion, neither the fact that sometimes through false
coincidences Alice might think that teleportation occurred nor the
fact that only one Bell state could be identified is relevant for
the teleportation fidelity.

\section{Quantum Teleportation of Entanglement}
\label{entangle} Our statement that the rather low efficiencies of
the first Innsbruck teleportation experiments do by no means
influence the fidelity is even more obvious in the second
experiment where, in a realisation of entanglement swapping, it
was possible to teleport a qubit which is still entangled to
another one. Figure 2 indicates the entanglement swapping
procedure and Fig. 4 is a schematic drawing of the experimental
setup. The main difference to the first experiment simply was that
photon 1, who's polarization properties had to be teleported, was
not prepared in a well-defined state prior to teleportation, but
rather in a measurement on its twin, photon 4, at a time after the
Bell state analyzer had registered a coincidence.  This,
undoubtedly, realises teleportation in a clear quantum situation,
since entanglement between two particles that did not share a
common origin nor interacted with one another in the past is the
very result of the teleportation procedure.

As in the first experiment, here too one has to deal with the case
that Alice might have false coincidence counts at her Bell state
analyzer together with a count at detector D$_4$ for photon 4 (see
Fig. 4). This again simply indicates that  two pairs have been
emitted to the left with no photon going to Bob. As above, since
it is intended to teleport only single-photon qubits, it is an
advantage that teleportation did not occur in this case.

In the entanglement teleportation experiment a linear polarizer in
front of the detector of photon 4 is set at various angles
($\Theta$). As a consequence, whenever teleportation succeeds, the
photon received by Bob should be orthogonal to the detected
polarization state of photon 4 (since both pairs 1 and 4, and 2
and 3 are prepared in the anti-symmetric state $\Psi^-$, and since
this state is also monitored by Alice, photon 3 and 4 will be
entangled in exactly this state, too (see section 1 and
Ref.~\cite{PAN98}).

This can be verified by performing a polarization measurement on
photon 3 carrying the teleported polarization properties of photon
1. In the experiment it was decided to register the  coincidences
between the two polarization measurements on photons 3 and 4 as a
function of the relative angle between the two polarizations (the
polarization of photon 3 is measured in the
 +45$^{\circ}$/-45$^{\circ}$ basis using a $\lambda/2$
rotation plate and a polarizing beamsplitter, while the
polarization of photon 4 is measured after passing a variable
polarizer at angle $\Theta$). This is equivalent to a measurement
of two-qubit correlations in a Bell inequality experiment
(\cite{ASPECT,WEISH}).

Again, since Alice identified one Bell state only, the coincidence
rate is reduced. Yet, clearly, the observed coincidence counts
show correlations well above the classical maximum of 50\%
visibility and will violate a Bell-type inequality as soon as the
coincidence fringe visibility surpasses the critical threshold
value of 71\%\cite{BELL}. This visibility was actually surpassed
in an individual run of the experiment where alignment and
stability parameters appear to have been very favorable. This, and
the regular visibility of (65 $\pm$ 2)\% (Fig. 5., corresponding
to a fidelity of 0.82 $\pm$ 0.01) indicate that it will be
possible to actually demonstrate a violation of Bell's inequality
in the near future.

In the case of entanglement teleportation it is really obvious
that it is wrong to use a Fock-state description and to include
the vacuum state for those cases where teleportation did not occur
in the definition of the teleportation fidelity, as has been
suggested by Braunstein and Kimble. To underline their claim, they
suggested that, instead of following the teleportation protocol as
described above, Bob could simply use randomly polarized photons
to obtain the same (or even better) teleportation fidelity. Yet,
clearly, if Bob were to follow that procedure, it would never be
possible to observe non-classical correlations and to achieve a
violation of a Bell-type inequality. Indeed, the observed
coincidence count rates (D$^-_3$D$_4$ and D$^+_3$D$_4$) would not
even show the sinusoidal variation as function of $\Theta$
exhibited in Fig. 5.

\section{Concluding Remarks}
\label{conc} In this contribution we demonstrated explicitly the
high fidelity, of the order of 0.8, achieved in the teleportation
experiments first performed in Innsbruck.  The measurement of the
fidelity of the teleportation is based on a four-fold coincidence
detection technique. The detection of Bob's photon (photon 3 in
Fig. 1) plays the double role of projecting out onto the
single-photon input state and of measuring the overlap of the
single-photon input state with the teleported single-photon state.
The role of projecting onto a single-photon input state can be
omitted if other means of preparing a single photon input state
had been used. This is however a technical, though difficult,
issue that has nothing to do with the actual quantum teleportation
procedure and therefore the teleportation fidelity will be exactly
the same in such situations.

Even if, more or less for technical reasons, the efficiency of the
experiments discussed above was very low, the data shown cannot be
obtained with any classical communication procedure. Moreover,
they clearly demonstrate the capability of this teleportation
procedure being implemented as a quantum channel for other quantum
communication schemes, e.g., for quantum cryptography, and that
the bona fide receiver can be quite sure about the fidelity of the
teleported qubit.

The fact that the discussion about the Innsbruck experiments has
not abated yet gives us the impression that our initial reply
(\cite{COM2}) to the criticism voiced by Braunstein and Kimble
(\cite{COM1}) might have been too succinct and condensed. We hope
that our present paper will help clarify the essential points such
that the debate can be set to rest.

\section*
{acknowledgment}
This work was supported by the Austrian Science Foundation
FWF Project Nos S6502 and F1506, the Austrian Academy of Sciences and TMR program
of the European Union (Network contract No. ERBFMRXCT96-0087).

\newpage

%FIG.1%

\begin{figure}
\caption{Experimental set up for the teleportation of qubits:
 A pulse of ultraviolet (UV) light passing
through a non-linear crystal can create the ancillary pair of
entangled photons 2 and 3. After retroflection during its second
passage through the crystal the ultraviolet pulse can create another
pair of photons, one of which will be prepared in the initial state
of photon 1 to be teleported, the other one serves as a trigger
indicating that a photon to be teleported is under way. Alice then
looks for coincidences after a beam splitter (BS) where the initial
photon and one of the ancillaries are superposed. Bob, after
receiving the classical information that Alice obtained a
coincidence count in detectors f1 and f2 identifying  the
$|\Psi^{-} \rangle_{12}$ Bell-state, knows that his photon 3 is in
the initial state of photon 1 which he then can check using
polarization analysis with the polarizing beam splitter (PBS) and
the detectors d1 and d2. The detector P provides the information
that photon 1 is under way.}
\end{figure}

%FIG.2%
\begin{figure}
\caption{Principle of teleportation of entanglement, also known as
entanglement swapping: Two EPR sources produce two pairs of
entangled photons, pair 1-4 and pair 2-3. Two photons, one from
each pair (photons 1 and 2) are subjected to a Bell-state
measurement(BSM). This results in projecting the other two
outgoing photons 3 and 4 onto an entangled state.}
\end{figure}

%FIG.3%
\begin{figure}
\vspace{2cm} \caption{Fidelity of teleportation of a qubit encoded
in the polarisation of a single-photon state: The overlap of the
input qubit (represented by photons linear polarised along (a)
45$^{\circ}$ and (b) 90$^{\circ}$) with the teleported qubit has
been determined via a four-fold coincidence technique to be as
high as 80\%. This very clearly surpasses the limit of 2/3,
indicated by the dotted lines, which at best could have been
obtained if Alice and Bob had been restricted to classical
communication only. }
\end{figure}

%FIG.4%
\begin{figure}
\vspace{2cm} \caption{Experimental setup for entanglement
swapping, i.e. teleportation of entanglement: A UV-pulse passing
through a non-linear crystal can create pair 2-3 of entangled
photons. Photon 2 is directed to the beamsplitter (BS).  After
reflection, during its second passage through the crystal the
UV-pulse can creates a second pair 1-4 of entangled photons.
Photon 1 will also be directed to the beamsplitter to perform a
Bell-state measurement (BSM) of photons 1 and 2. When photons 1
and 2 yield a coincidence click on the two detectors behind the
beamsplitter a projecting onto the $\left|\Psi^-
\right\rangle_{12}$ state takes place. As a consequence photons 3
and 4 will also be projected onto an entangled state. To analyse
their entanglement we look at coincidences between detectors
D$3^+$ and D4, and between detectors D$3^-$ and D4, for different
polarization angles $\Theta$.}
\end{figure}

%FIG.5%
\begin{figure}
\vspace{2cm}
\caption{Entanglement verification:
Four-fold coincidences, resulting from
two-fold coincidence D$3^+$D4 and D$3^-$D4 conditioned on the
two-fold coincidences at the Bell state measurement,
as function of the polarization angle $\Theta$.
The two complementary sine curves with a visibility of
$0.65\pm0.02$ demonstrate that photons 3 and 4
are polarisation entangled.}
\end{figure}

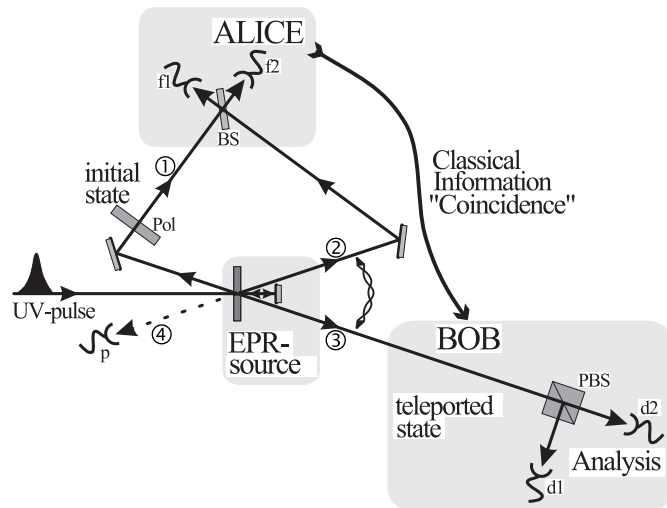
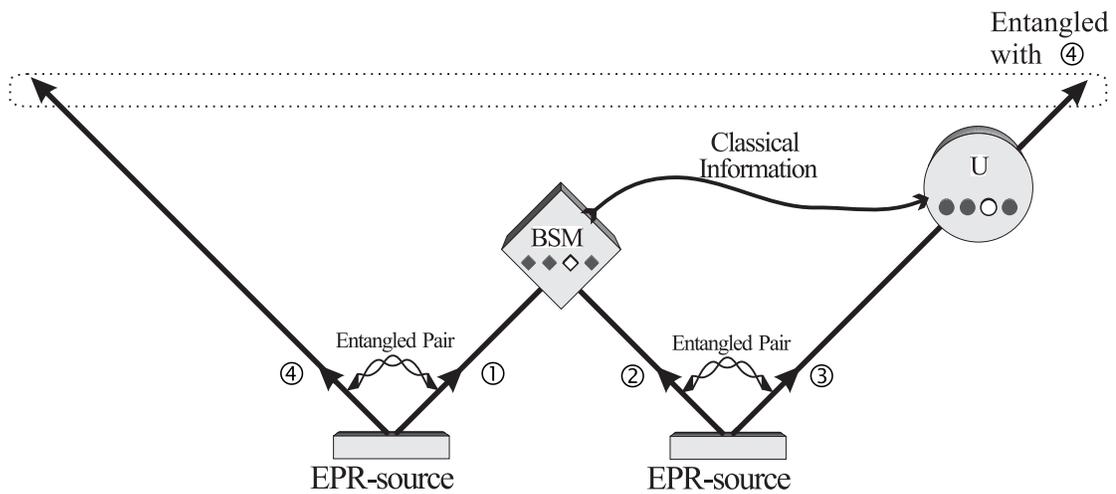
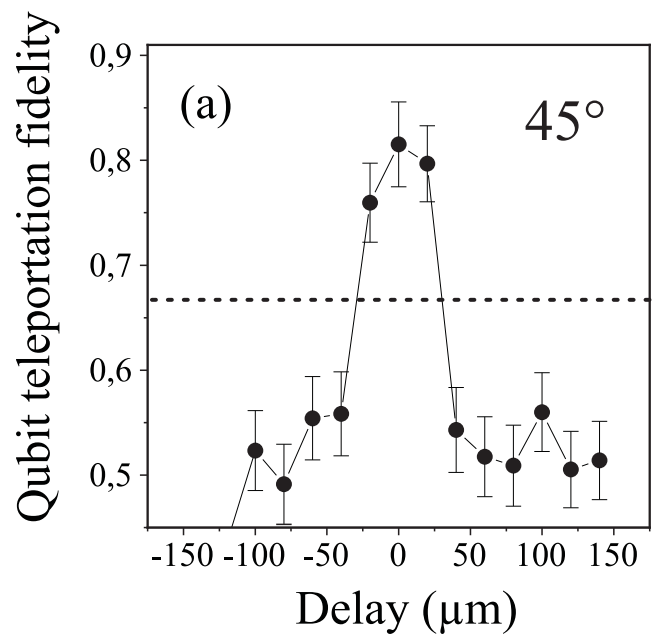
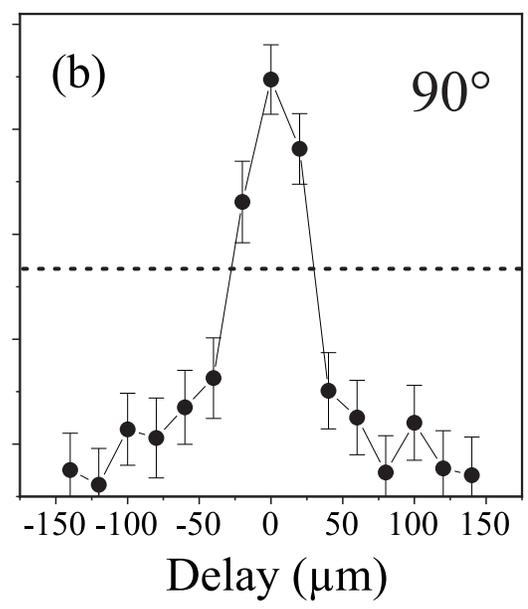
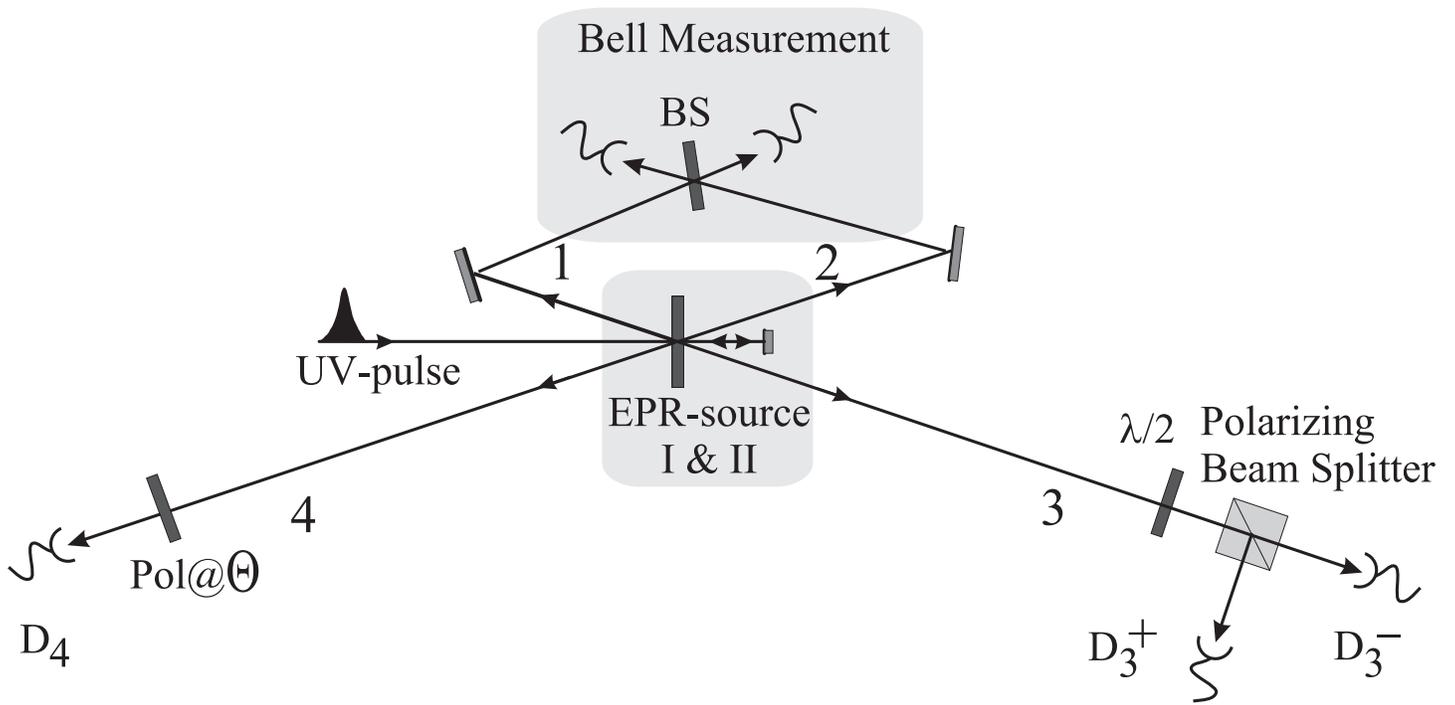
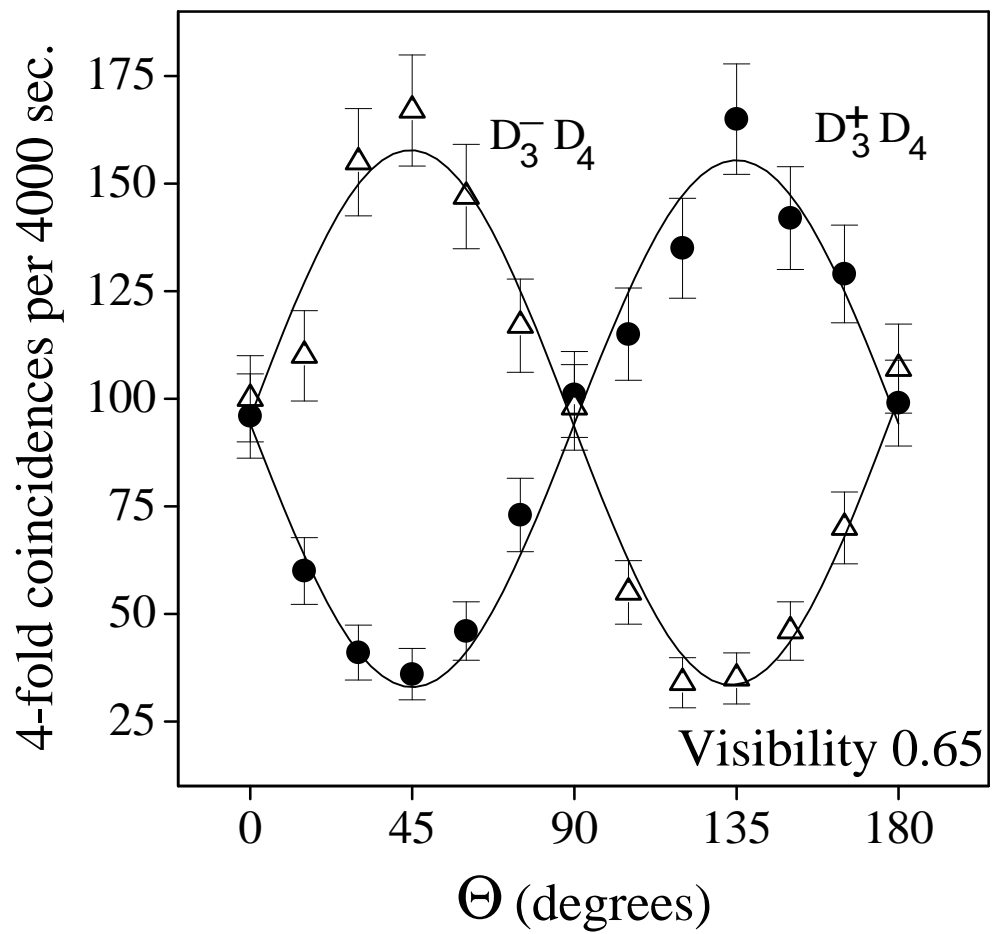
\end{document}